\documentstyle[floats,aps,epsf,prl]{revtex}

\begin{document}
\draft
\twocolumn[\hsize\textwidth\columnwidth\hsize\csname @twocolumnfalse\endcsname

\title{Integer quantum Hall effect of interacting electrons: dynamical
  scaling and critical conductivity}

\author{Bodo Huckestein and Michael Backhaus}

\address{Institut f\"ur Theoretische Physik, Universit\"at zu K\"oln,
  D-50937 K\"oln, Germany}

\date{\today}

\maketitle

\begin{abstract}
  
  We report on a study of interaction effects on the polarization of
  a disordered two-dimensional electron system in a strong magnetic
  field. Treating the Coulomb interaction within the time-dependent
  Hartree-Fock approximation we find numerical evidence for dynamical
  scaling with a dynamical critical exponent $z=1$ at the integer
  quantum Hall plateau transition in the lowest Landau level. Within
  the numerical accuracy of our data the conductivity at the
  transition and the anomalous diffusion exponent are given by the
  values for non-interacting electrons, independent of the strength of
  the interaction.

\end{abstract}
\pacs{PACS: 73.40.Hm, 71.30.+h, 72.15.Rn}
\vskip2pc

]

While the occurrence of quantized plateaus in the Hall conductivity of
two-dimensional systems in strong magnetic fields is the most striking
aspect of the quantum Hall effect (QHE) \cite{KDP80}, it is the
transitions between these plateaus that have recently attracted a lot
of attention, both experimentally and theoretically
\cite{reviews}. Our understanding of the QHE is not
complete without an understanding of the plateau transitions.
Furthermore, the plateau transitions are among the most extensively
studied examples of quantum phase transitions. The origin of the
plateaus is the localization of the electrons due to disorder, with
the plateau transitions corresponding to the
localization-delocalization transitions in the disorder potential
\cite{reviews}. Theoretical studies have focused on the effect of
disorder on non-interacting electrons in a strong magnetic field. From
numerical and analytical calculations the following picture emerged
for weak disorder\cite{reviews}: in every Landau level
exists a critical energy $E_c$ at which the localization length $\xi$
diverges as a power law $|E-E_c|^{-\nu}$ with a localization length
exponent $\nu=2.35\pm0.03$ independent of Landau level index and
correlation length of the disorder potential \cite{Huc}. At all
other energies electrons are exponentially localized. In the region of
localized states the longitudinal conductivity vanishes and the Hall
conductivity is quantized in multiples of $e^2/h$. At the critical
energies the Hall conductivity jumps by $e^2/h$ and the longitudinal
conductivity takes on a finite value $\sigma_c$ with
$\sigma_c\approx0.5e^2/h$ in the lowest Landau level
\cite{CD88,HHB93}. The eigenstates at the critical energies show
multifractal fluctuations leading to correlation functions being
characterized by an anomalous diffusion exponent $\eta\approx0.4$
\cite{CD88}. Dynamical correlation functions show scaling behavior as a
function of the variable $qL_\omega$, where $q$ is the wave-vector and
$L_\omega = 1/\sqrt{\rho_0\hbar\omega}$ is the relevant
frequency-dependent length scale, the effective system size with mean
level spacing $\hbar\omega$ \cite{CD88}. Comparing this result to the
definition of the dynamical critical exponent
$L_\omega\propto\omega^{-1/z}$ and observing that the density of states
(DOS) $\rho_0$ is finite at the transition, the dynamical critical
exponent is found to be $z=2$.

Experimentally, strong evidence for scaling behavior near the plateau
transitions was observed \cite{reviews}. In particular,
the observed values for the localization length exponent
$\nu=2.3\pm0.1$ \cite{KHKP91b} and the critical conductivity
$\sigma_c\approx0.5e^2/h$ \cite{Sea95} are in remarkable agreement
with the theory for non-interacting electrons \cite{Sachdev}. However, in
high-frequency measurements scaling with a dynamical critical exponent
$z=1$ was observed in contrast to the value $z=2$ for non-interacting
electrons \cite{ESKT93}. This discrepancy has been attributed to the
influence of the Coulomb interactions between the electrons that are
always present in the experiments \cite{LW96}. In fact, numerical
simulations in which the interactions are treated in self-consistent
Hartree-Fock (HF) approximation show a linear suppression of the DOS
at the Fermi energy irrespective of the position of the Fermi energy
\cite{YM93,YMH95}.  According to the argument presented above at the
critical energy this non-critical effect leads to a reduction of the
dynamical critical exponent $z$ from 2 to 1. The influence of the
Coulomb interactions on other critical properties of the QH
transitions was also studied within the self-consistent HF
approximation \cite{YMH95}. From the scaling of the participation
ratio of the HF eigenstates it was found that the localization length
exponent $\nu$ and the fractal dimension $D_2=2-\eta$ were not changed
by the interactions. In lieu of the critical conductivity the Thouless
numbers of the HF eigenenergies were studied and also found to be
unchanged by the interactions.  However, not much is known about the
relation of the conductivity of an interacting electron system and the
Thouless numbers of the HF eigenvalues.

In this paper we overcome some of the deficiencies of the previous
studies by directly calculating the irreducible dynamical polarization
$\Pi^{\text{irr}}(q,\omega)$. This allows us to obtain the critical
quantities $z$, $\sigma_c$, and $\eta$ from the scaling limit of the
polarization. Let us first sketch our approach. As the starting point
for our numerical calculations we use the self-consistent HF
approximation \cite{YM93,YMH95}. The corresponding conserving
approximation for the two-particle Green's function is the
time-dependent Hartree-Fock (TDHF) approximation \cite{KB62}. Within
this approximation we evaluated the dynamical polarization
$\Pi^{\text{irr}}(q,\omega)$ including vertex corrections. The
disorder-averaged polarization is extrapolated to infinite system size
and vanishing $q$ and $\omega$. In this limit, the
polarization obeys the scaling form
\begin{equation}
  \label{scale}
  \Pi^{\text{irr}}(q,\omega) =
  \frac{\sigma^*q^2/e^2}{\sigma^*q^2/e^2\chi^{\text{irr}}_q - i\omega}.
\end{equation}
The Onsager conductivity $\sigma^*$ is related by the Einstein
relation
\begin{equation}
  \label{einstein}
  \sigma^* = e^2 \chi^{\text{irr}}_q D
\end{equation}
to the diffusion coefficient $D(q,\omega)$ and the static
susceptibility $\chi^{\text{irr}}_q$. For vanishing wave-vector it
coincides with the Kubo conductivity $\sigma =
e^2\omega\Pi^{\text{irr}}/iq^2$. Due to the occurrence of the Coulomb gap
even at criticality the susceptibility $\chi^{\text{irr}}_q\propto q$.
The conductivity $\sigma^*(q,\omega)$ is found to be a function of the
scaling variable $x=q^2/\chi^{\text{irr}}_q\hbar\omega$ with the limiting
behavior
\begin{equation}
  \label{sigma_limit}
  \sigma^*(x) \longrightarrow \left\{ 
      \begin{array}{cl}
        \sigma^*_c, & x \rightarrow 0,\\
        \propto x^{1-\eta}, & x\rightarrow \infty.
      \end{array}
      \right.
\end{equation}
Since $x\propto q/\omega$ the Coulomb gap changes the dynamical
critical exponent $z$ from its non-interacting value 2 to 1.  On the
other hand, we find that the critical DC conductivity
$\sigma^*_c=0.5\pm0.1e^2/h$ and the anomalous diffusion exponent
$\eta=0.4\pm0.1$ are independent of the strength of the interactions
and in agreement with previous calculations for non-interacting
electrons \cite{CD88,HHB93,HS94}. These are the central results of our 
study.

In the remainder of the paper we present our model, our calculations
and our results, relegating a more detailed discussion to a separate
publication \cite{BH98a}. We consider the situation where the
cyclotron energy $\hbar\omega_c$ is much larger than both the disorder
and the Coulomb energy and the Fermi energy is in the lowest Landau
level onto which we project the Hamiltonian. As a model for the
disorder we use the random Landau matrix \cite{HK89}.  The
strength of the disorder potential is characterized by the level
broadening in self-consistent Born approximation $\Gamma$ and its
range is taken to be zero. The electrons interact via the bare Coulomb
interaction $V(q)=2\pi e^2/\kappa q$. The strength of the interactions
is given by the ratio $\gamma=(e^2/\kappa l)/\Gamma$, where $\kappa$
is the dielectric constant and $l$ is the magnetic length. The square
systems of linear size $L$ with periodic boundary conditions contained
$N$ electrons and from $N_\Phi=L^2/2\pi l^2=169$ to 900 flux quanta.
In the presence of interactions physical quantities depend on the
filling factor $\nu_F=N/N_\Phi$. We choose $\nu_F=1/2$ in all our
calculations so that the Fermi energy coincides with the critical
energy of the plateau transition.  Including the HF interaction the
eigenvalues $\epsilon_\alpha$ and eigenfunctions $|\alpha\rangle$ of
the Hamiltonian are calculated self-consistently \cite{YM93,YMH95}. A
major simplification arises in this calculation since in the lowest
Landau level the exchange interaction
\begin{equation}
  \label{Vex}
  V_{\text{ex}}(q,q') =
  -\delta_{q,-q'} \sqrt{2\pi^3}e^2l\kappa^{-1} \text{e}^{q^2l^2/4}
  \text{I}_0(q^2l^2/4)
\end{equation}
and hence the HF interaction
\begin{equation}
  \label{VHF}
    V_{\text{HF}}(q,q') =
  \delta_{q,-q'} \left(V(q)+V_{\text{ex}}(q)\right)
\end{equation}
become local ($\text{I}_0$ is the Bessel function) \cite{FPA79,MG88}.

\begin{figure}
  \begin{center}
    \epsfxsize=6.4cm
    \leavevmode
    \epsffile{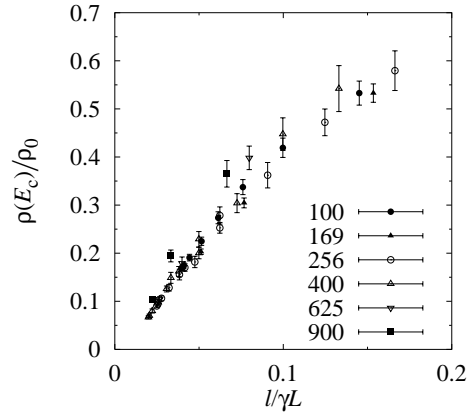}
    \caption{Single-particle density of states at the critical energy
      as a function of $l/\gamma L$ for $N_\Phi$ between 100 and 900
      and $\gamma$ between 0.15 and 1.}
    \label{fig:DOS}
  \end{center}
\end{figure}

Perhaps the most characteristic feature of the HF spectrum is the
formation of a Coulomb gap in the single-particle DOS \cite{ES75}. In
particular, the DOS at the Fermi energy $\rho(E_F)$ decreases linearly with
$l/\gamma L$. This is shown in Fig.~\ref{fig:DOS} \cite{YM93,YMH95}.
The suppression of the DOS is not related to the phase transition but
occurs for all values of the Fermi energy. Only when the Fermi energy
coincides with the critical energy does it change the critical
dynamics of the system.

The bare polarization $\Pi^0({\bf q},{\bf q}',\omega)$ is given by
\begin{eqnarray}
  \label{Pi_0}
  \lefteqn{\Pi^0({\bf q},{\bf q}',\omega)=\sum_{\alpha,\beta}
  (\alpha|\beta)_{\bf q} (\beta|\alpha)_{{\bf q}'}} \nonumber\\
  && \times\left[
    \frac{f_\alpha(1-f_\beta)}{\hbar\omega+\epsilon_\beta-\epsilon_\alpha-i\delta}
    -
    \frac{f_\beta(1-f_\alpha)}{\hbar\omega+\epsilon_\beta-\epsilon_\alpha+i\delta}
    \right],
\end{eqnarray}
where $(\beta|\alpha)_{\bf q} =
\langle\beta|\text{e}^{-i{\bf qr}}|\alpha\rangle$, $f_\alpha$ is
the Fermi function (we take $T=0$ in our calculations) and $\delta$ is
an infinitesimal parameter of the order of the mean level
spacing. After averaging over disorder (we use between 13 and
100 realizations of the disorder) $\Pi^0$ becomes translationally and
rotationally invariant $\Pi^0({\bf q},{\bf q}',\omega) =
\delta({\bf q}+{\bf q}')\Pi^0(q,\omega)$. The real part of the static
bare polarization coincides in the limit $q\to0$ with the single
particle DOS, $\lim_{q\to0}\text{Re}\Pi^0(q,\omega=0) = \chi^0_0 =
\rho(E_F)$ \cite{BH98a}.

\begin{figure}
  \begin{center}
    \epsfxsize=6.4cm
    \leavevmode
    \epsffile{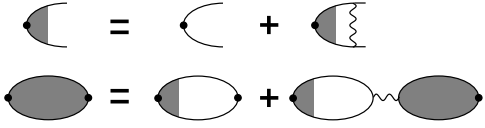}
    \caption{Feynman diagrams for the irreducible vertex (top) and the
      screened polarization (bottom) in time-dependent Hartree-Fock
      approximation.}
    \label{fig:feyn}
  \end{center}
\end{figure}

The TDHF approximation for the irreducible and the screened
polarizations are shown in Fig.~\ref{fig:feyn}.
In the lowest Landau level the ladder sum for the vertex corrections
can be transformed into a RPA-like bubble sum with the Coulomb
potential replaced by $V_{\text{ex}}(q)$ \cite{AM91,BH98a}. Similarly,
the screened polarization $\Pi^{\text{scr}}$ is related to the bare
polarization through a bubble sum containing the HF potential
$V_{\text{HF}}(q)$ \cite{BH98a},
\begin{mathletters}
  \label{Pis}
\begin{eqnarray}
  \Pi^{\text{irr}}({\bf q},\omega) &=& \Pi^0({\bf q},\omega) \left( 1
    - V_{\text{ex}}({\bf q})\Pi^{\text{irr}}({\bf q},\omega) \right), \\
  \Pi^{\text{scr}}({\bf q},\omega) &=& \Pi^0({\bf q},\omega) \left( 1
    - V_{\text{HF}}({\bf q})\Pi^{\text{scr}}({\bf q},\omega) \right).
\end{eqnarray}
\end{mathletters}
In principle, the vertex corrections should be applied to $\Pi^0({\bf
  q},{\bf q}')$ since it becomes diagonal only after disorder
averaging. We checked for small system sizes that using
Eq.~(\ref{Pis}) only introduces errors comparable to the statistical
uncertainties for the relevant parameter range.

Since the approximations leading to $\Pi^{\text{irr}}$ and
$\Pi^{\text{scr}}$ are conserving and due to Eq.~(\ref{Pis}), the
polarizations can be parameterized in the limit of infinite system size
and vanishing $q$ and $\omega$ through a susceptibility $\chi_q$ and a 
diffusion coefficient $D(q,\omega)$,
\begin{equation}
  \label{Pi_scale}
  \Pi^A(q,\omega) = \frac{\chi^A_q D^A q^2}{D^A q^2 - i\omega},
\end{equation}
with $A=\{0,\text{irr},\text{scr}\}$. The Einstein relation
$\sigma^A(q,\omega)=e^2\chi^A_q D^A(q,\omega)$ defines the Onsager
conductivity $\sigma^*$. From Eq.~(\ref{Pis}) it is apparent that 
the dressing of $\chi_q$ and $D$ leaves the Onsager conductivity
unchanged
\begin{equation}
  \label{sigmas}
  \sigma^* = e^2\chi^0_q D^0 = e^2\chi^{\text{irr}}_q D^{\text{irr}} =
  e^2\chi^{\text{scr}}_q D^{\text{scr}}.
\end{equation}
For vanishing wave-vector the Onsager conductivity coincides with the
Kubo conductivity $\sigma^*(q=0,\omega) = \sigma(\omega) = \lim_{q\to0}
e^2\omega\Pi^{\text{irr}}(q,\omega)/iq^2$.  

\begin{figure}
  \begin{center}
    \epsfxsize=6.4cm
    \leavevmode
    \epsffile{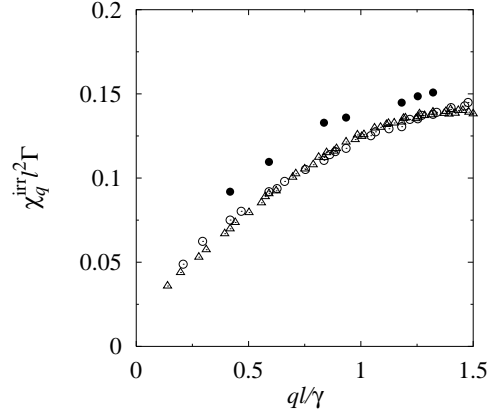}
    \caption{Static irreducible susceptibility for $N_\Phi=900$ and
      $\gamma=0.2$, 0.4 and 0.6. The finite intercept at $q=0$ scales
      like $l/L$ to zero.}
    \label{fig:chiq}
  \end{center}
\end{figure}

The polarization can be rewritten in terms of $\sigma^*$,
\begin{equation}
  \label{Pi_sigma}
  \Pi^{\text{irr}}(q,\omega) = \chi^{\text{irr}}_q \frac{\sigma^*
    x}{\sigma^* x - ie^2/\hbar},
\end{equation}
with $x = q^2/\chi^{\text{irr}}_q\hbar\omega$. If dynamical scaling holds
then $\sigma^* x$ is a function of $q^z/\omega$. In order to
extract the scaling behavior from our numerical data for finite system
sizes, we first calculate the static susceptibility
$\chi^{\text{irr}}_q$ from the real part of the irreducible
polarization. Due to the occurrence of the linear Coulomb gap (cf.
Fig.~\ref{fig:DOS}), $\chi^{\text{irr}}_q\propto q$ for small $q$ in
the thermodynamic limit, as is shown in Fig.~\ref{fig:chiq}.

\begin{figure}
  \begin{center}
    \epsfxsize=6.4cm
    \leavevmode
    \epsffile{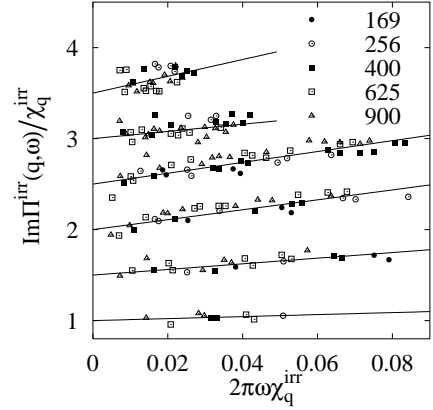}
    \caption{Extrapolation of
      $\text{Im}\Pi^{\text{irr}}(q,\omega)/\chi^{\text{irr}}_q$ as
      described in the text. $x/2\pi=0.5$, 1, 1.5, 2, 4, 6 (bottom to
      top).}
    \label{fig:extra}
  \end{center}
\end{figure}

Next, the imaginary part of $\Pi^{\text{irr}}$ is divided by the
numerically obtained $\chi^{\text{irr}}_q$. The resulting function is
then simultaneously extrapolated to infinite system size and $q$ and
$\omega$ to zero for fixed value of $x$ as is shown in
Fig.~\ref{fig:extra} \cite{CD88,BH98a}. This choice 
of scaling variable allows for reasonable extrapolations, while other
choices like $q^2/\rho_0\hbar\omega$ do not give satisfactory results. 
Since $\chi^{\text{irr}}$ is asymptotically linear in $q$, $x\propto
q/\omega$ in the thermodynamic limit and the dynamical critical
exponent $z=1$.

\begin{figure}
  \begin{center}
    \epsfxsize=6.4cm
    \leavevmode
    \epsffile{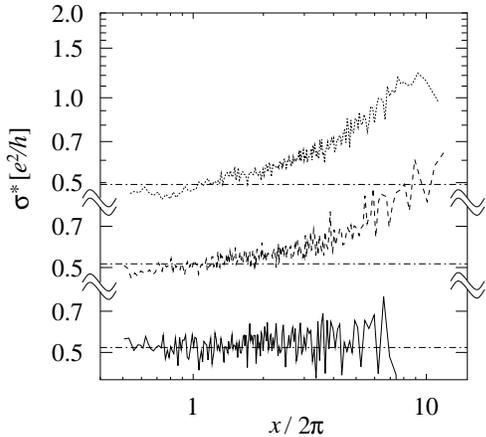}
    \caption{Conductivity $\sigma^*$ as a function of the scaling
      variable $x=q^2/\chi^{\text{irr}}_q\hbar\omega$ for
      $\gamma=0.2$, 0.4 and 0.6 (bottom to top). The curves are
      shifted vertically.}
    \label{fig:sigma}
  \end{center}
\end{figure}

Figure \ref{fig:sigma} shows the resulting conductivity $\sigma^*$ as
a function of $x$.
In the limit $x\to0$, $\sigma^*$ coincides with the DC Kubo
conductivity $\sigma_c$. We find that $\sigma_c$ does not depend on
the strength of the interaction up to $\gamma=0.6$ and is given by
$\sigma_c=0.5\pm e^2/h$. In the opposite limit $x\to\infty$, the
conductivity reflects multifractal density fluctuations and scales
like $x^{1-\eta}$ \cite{BH98a}. A numerically more accurate
determination of $\eta$ is possible from the scaling of the static
polarization \cite{BH98a}. We find $\eta=0.4\pm0.1$ independent of the 
interactions and in agreement with published data for non-interacting
electrons \cite{CD88,HS94}. Comparing our result to \cite{YMH95}
we see that the irreducible polarization reflects the multifractal
exponent $D_2$ of the HF eigenstates similar to the case of
non-interacting electrons.

Let us make a few remarks on our results. Although the conductivity
$\sigma_c$ is finite at the transition, the vanishing
susceptibility $\chi^{\text{irr}}_q$ implies non-diffusive behavior of
the HF-particles, $D(q,\omega)\propto q^{-1}$ for $q\to0$. A
central issue is the validity of the HF approach that has recently
been questioned \cite{PS98}. The change in the dynamical
critical exponent is due to the use of the unscreened Coulomb
interaction. The dielectric function $\kappa$ is given by
$\kappa(q,\omega) = 1 + V(q) \Pi^{\text{irr}}(q,\omega)$. From
eq.~(\ref{Pi_scale}) it follows that screening behaves differently in
the limits of small and large $x$. In the limit of small $x$, that is
relevant to transport, screening is inefficient and $\kappa\to
1$. The use of the unscreened HF approximation thus might be justified 
for the calculation of the frequency-dependent conductivity. On the
other hand, in the limit of large $x$ we observe static screening,
$\kappa(q,\omega)=1 + \chi^{\text{irr}}_qV(q)$. For slow processes,
during which the ground state has time to relax, the statically screened HF
approximation is more appropriate. It leads to a finite susceptibility 
and is described by $z=2$ \cite{BH98a}. A final judgment on the
validity of the HF approach to the critical quantum Hall system might
only by possible if one succeeds in a renormalization group treatment
of the neglected correlations \cite{LW96,Pruisken}.

In summary, we have presented results on the influence of Coulomb
interactions on the critical properties of the integer quantum Hall
system. Within the time-dependent Hartree-Fock approximation we find
the critical conductivity $\sigma_c$ and the anomalous diffusion
exponent $\eta$ are unchanged by interactions while the dynamical
critical exponent $z$ changes from 2 to 1. The origin of this behavior 
is the non-critical reduction of the single-particle density of
states.

This work was performed within the research program of the SFB 341 of
the DFG.


\end{document}